# Axions from String Decay

C. Hagmann[a], S. Chang[b], and P. Sikivie[b]

[a] Lawrence Livermore National Laboratory, 7000 East Ave, Livermore, CA 94550

[b] University of Florida, Physics Department, Gainesville, FL 32611

We have studied numerically the evolution and decay of axion strings. These global defects decay mainly by axion emission and thus contribute to the cosmological axion energy density. The relative importance of this source relative to misalignment production of axions depends on the spectrum. Radiation spectra for various string loop configurations are presented. They support the contention that the string decay contribution is of the same order of magnitude as the contribution from misalignment.

## 1. INTRODUCTION

Axion strings arise in the early universe when the global $U(1)_{PQ}$ symmetry breaks spontaneously at the scale $f_a \sim 10^{12}$ GeV. A Brownian network of strings is formed initially with O(1) string per horizon volume. As the horizon expands, wiggles on the strings will start oscillating and radiate axions. Strings will also intercommute and form closed loops, which collapse and convert into axions. Simulations of gauge string networks show that a scaling solution [1,2] is reached after a few Hubble times with about one long string per horizon size and a population of decaying loops. This process continues until QCD time, when the axion acquires a small mass. As a consequence, domain walls form between strings. The wall-string system is short-lived because wall tension pulls the strings together, followed by mutual annihilation into free axions. If there is an inflationary period after PQ symmetry breaking with a reheat temperature $< f_a$, strings get diluted and do not contribute to the cosmological axion density.

The dynamics of axion strings is governed by the classical Lagrangian density

$$L=\frac{1}{2}\partial_\mu\phi^*\partial^\mu\phi - \frac{1}{4}\lambda(\phi^*\phi-f_a^2)^2 \quad (1)$$

where $\phi = |\phi|\exp(ia/f_a)$ is a complex scalar field whose phase is the axion field. A simple solution of (1) is a static, straight global string with core size $\delta = (\sqrt{\lambda} f_a)^{-1}$ and energy per unit length

$$\mu \approx \pi f_a^2 \ln(L/\delta) \quad (2)$$

where $L$ is the long distance cutoff, *i.e.* the interstring distance. The logarithmic factor in Equation 2 distinguishes a global from a gauge (local) string, whose energy density decays exponentially outside of its core. For axion strings near QCD time, $\ln(t_{QCD} f_a) \approx 70$ and most of the string energy resides outside of the core.

The number of axions emitted by strings during the string epoch (1 GeV $< T < 10^{12}$ GeV) can in principle be calculated from the radiation spectrum of the evolving string network. Analytic techniques are not well developed and one must rely on numerical simulations. A major difficulty is the enormous loop to core size ratio of realistic axion strings. Currently, the largest computer simulations have $\ln(L/\delta) \approx 7$. Yet, it is valuable to study the spectrum as a function of $\ln(L/\delta)$ in order to deduce a possible trend. This is especially important for closed loops as they dissipate most of the string network energy.

In the past, there has been considerable controversy over the correct radiation spectrum. One group [3,4,5,6,7] (scenario A) argues that the spectrum is strongly peaked at wavelengths of order the loop size. This corresponds to an under-damped decay with about 5-10 oscillations per loop half-life when $\ln(L/\delta) \approx 70$. A second group [8,9,10] (scenario B) argues that the loops decay without oscillations, in a time of order the initial size divided by the speed of light. The radiation spectrum is $dE/dk \sim 1/k$ (which is also the spectrum of the static string field) with cutoffs at the inverse loop size and core thickness. Of course, it is possible that neither scenario A or B is correct.

The cosmological axion number density at time $t$ is related to the radiation spectrum through

$$n_a(t) \sim \frac{1}{t^{3/2}} \int_{t_{PQ}}^{t} \frac{dt'}{t'^{3/2}} \frac{\mu(t')}{\overline{\omega}(t')} \tag{3}$$

where

$$\frac{1}{\overline{\omega}(t)} = \int \frac{dE}{d\omega}(t) \frac{d\omega}{\omega} \Big/ \int \frac{dE}{d\omega}(t)\, d\omega \tag{4}$$

and $dE/d\omega$ is the radiated axion energy spectrum. A useful quantity is

$$N_{ax}(t) = \int \frac{dE}{dk}(t) \frac{dk}{k}. \tag{5}$$

In scenario B, $N_{ax}$ is predicted to stay constant since both axion radiation and static string field have the same spectrum. Conversely, one expects $N_{ax}$ to increase by $\ln(L/\delta)$ in scenario A. The factor $r = N_{ax}^{\text{final}} / N_{ax}^{\text{initial}}$, by which $N_{ax}$ increases during the decay of a string loop determines the string decay contribution to the axion cosmological energy density. The string decay contribution is $r$ times the contribution from misalignment. In scenario A, $r$ is of the order $\ln(L/\delta)$, whereas in scenario B, $r$ is of order unity.

## 2. LOOP SIMULATIONS

We have performed simulations of various loop geometries: (1) circular loops initially at rest, (2) noncircular loops with angular momentum, and (3) string-antistring pairs with angular momentum. The initial configurations are set up on large (~$10^7$ points) Cartesian grids, and then time-evolved using the finite-difference equations derived from Eq. (1). A FFT spectrum analysis of the kinetic and gradient energies during the collapse yields $N_{ax}(t)$.

### 2.1. Circular loops

Because of azimuthal symmetry, circular loops can be studied in r-z space. By mirror symmetry, the problem can be further reduced to one quarter-plane.

The static axion field far from the string core is (with $f_a \equiv 1$)

$$a(r,z) \approx \frac{1}{2}\Omega(r,z) \tag{6}$$

in the infinite volume limit, where $\Omega$ is the solid angle subtended by the loop. We use as initial configuration the outcome of a relaxation routine starting with Equation 6 outside the core and

$$\phi(\rho) \approx 0.58(\rho/\delta)\exp(i\theta) \tag{7}$$

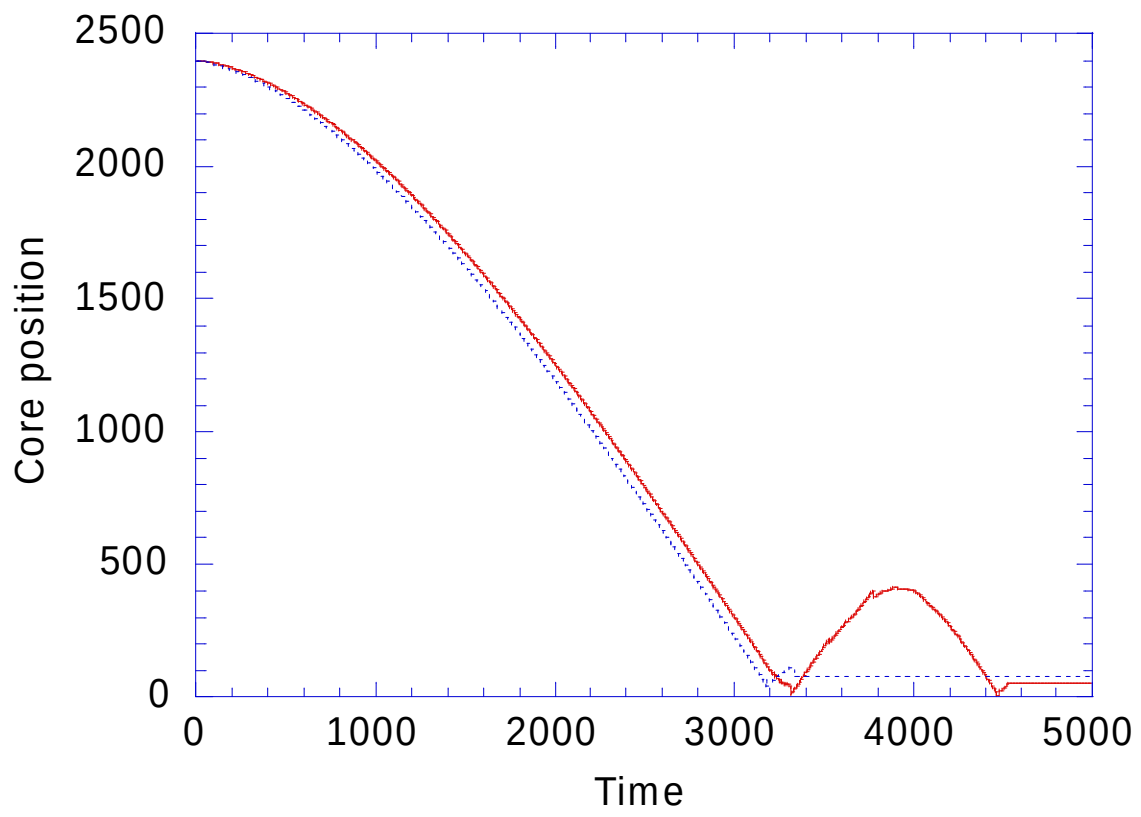

Figure 1: Core position of collapsing loop versus time for $\lambda$ = 0.001 (dotted line) and $\lambda$ = 0.004 (solid line). The lattice size is 4096×4096.

within the core. Here, $\rho$ is the distance to the string, and $\theta$ is the winding angle. The relaxation and the subsequent dynamical evolution are done with reflective boundary conditions. A step size $dt$ = 0.2 was used for the time evolution and the total energy was conserved to better than 1 %. In general the loops collapsed at nearly the speed of light without a rebound. For a small range of parameters, $80 < R_0/\delta < 190$, where $R_0$ is the initial loop radius, we noticed a small bounce as shown in Figure 1.

There is a substantial Lorentz contraction of the string core as it collapses (see Figure 2). A lattice effect became evident when the reduced core size becomes comparable to the lattice spacing. This lattice effect consists of a "scraping" of the string core on the underlying grid, during which the kinetic energy of the string gets dissipated into high frequency axion radiation. We always choose $\lambda$ small enough to avoid this phenomenon.

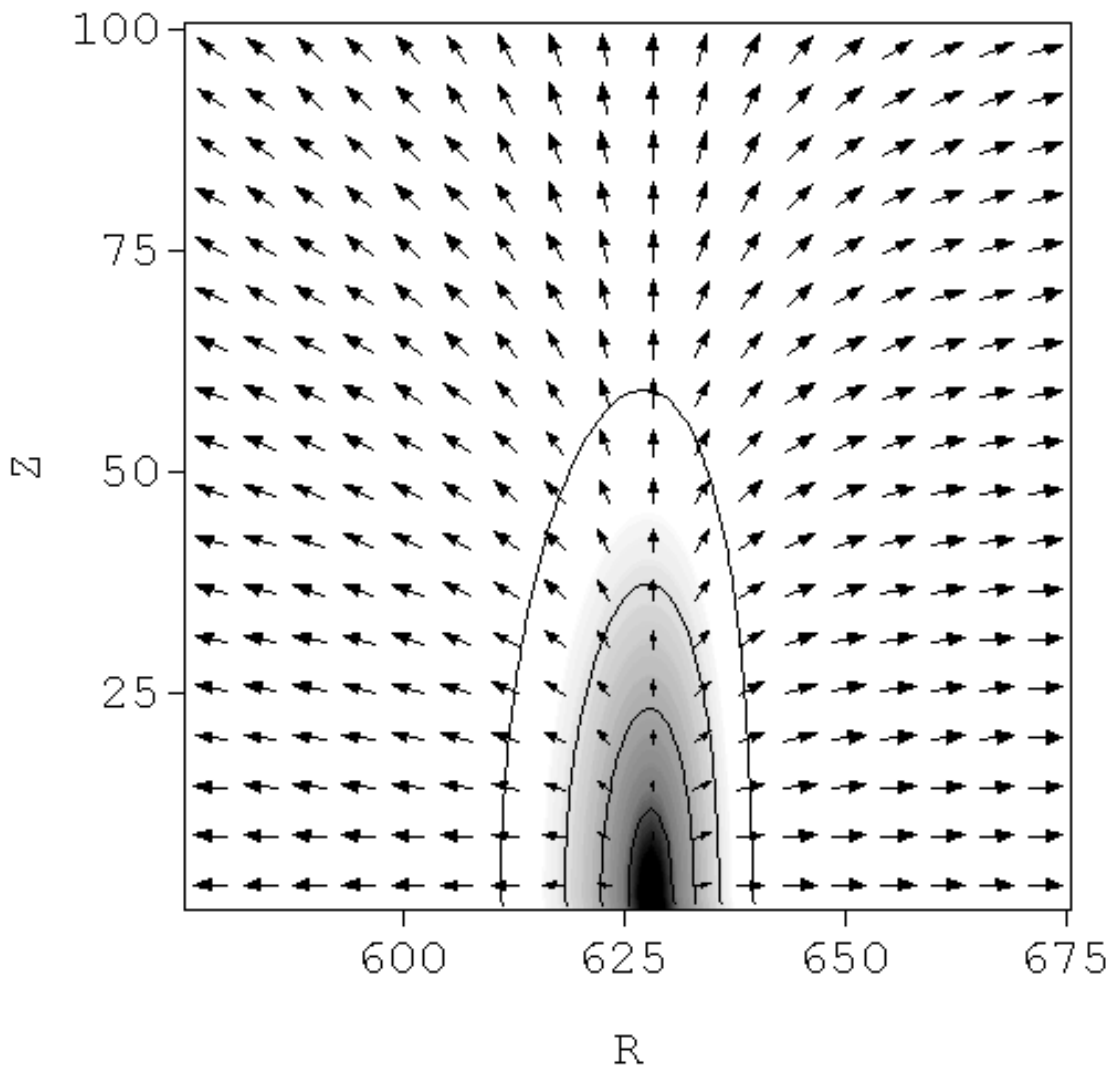

Figure 2: Intensity plot of potential energy (contours represent constant potential energy) in vicinity of string core for $R_0$ = 2400, λ = 0.001 at $t$ = 2600. The Lorentz factor is about 4. The arrows represent the axion field.

A spectrum analysis of the fields was performed by expanding the gradient and kinetic energies as

$$\begin{aligned}
\nabla_z \phi &= \sum_{mn} c_{mn}\, J_0(k_m r)\sin(k_n z) \\
\nabla_r \phi &= \sum_{mn} c_{mn}\, J_1(k_m r)\cos(k_n z) \qquad (8)\\
\dot{\phi} &= \sum_{mn} c_{mn}\, J_0(k_m r)\cos(k_n z)
\end{aligned}$$

with the boundary conditions $J_1(k_m R_{\max}) = 0$ and $\sin(k_n Z_{\max}) = 0$. The dispersion relationship is given by $\omega_{mn} = \sqrt{k_m^2 + k_n^2}$. Figure 3 shows the power spectrum $dE/d(\ln k)$ displayed in $\ln k$ bins of width $\Delta \ln k = 0.5$ at $t$ = 0 and after the collapse at $t$ = 3000. At both times, the spectrum exhibits an almost flat plateau, consistent with a $dE/dk \propto 1/k$ spectrum. The high frequency cutoff of the spectrum is increased however after the collapse and is associated with the Lorentz contraction of the core.

The evolution of $N_{ax} \equiv \sum_{mn}(E_{mn}/k_{mn})$ during the loop collapse was studied for various values of $R_0/\delta$. We observe a marked decrease of $N_{ax}$ by ~ 20 % during the collapse roughly independent of $R_0/\delta$.

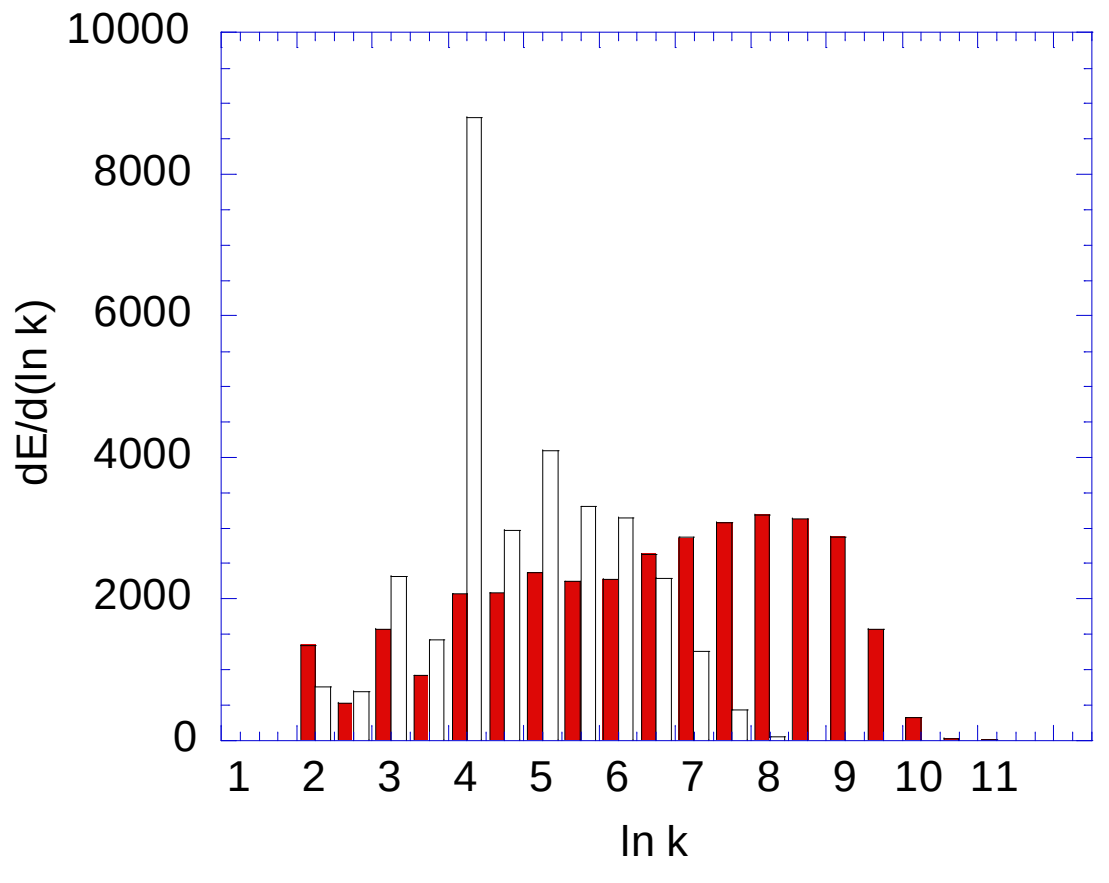

Figure 3: Energy spectrum of collapsing loop for $R_0$ = 2400 and λ = 0.001. The white (black) histogram represents the spectrum at $t$ = 0 ($t$ = 3000). The increased high frequency cutoff of the final spectrum is due to the Lorentz contraction of the core.

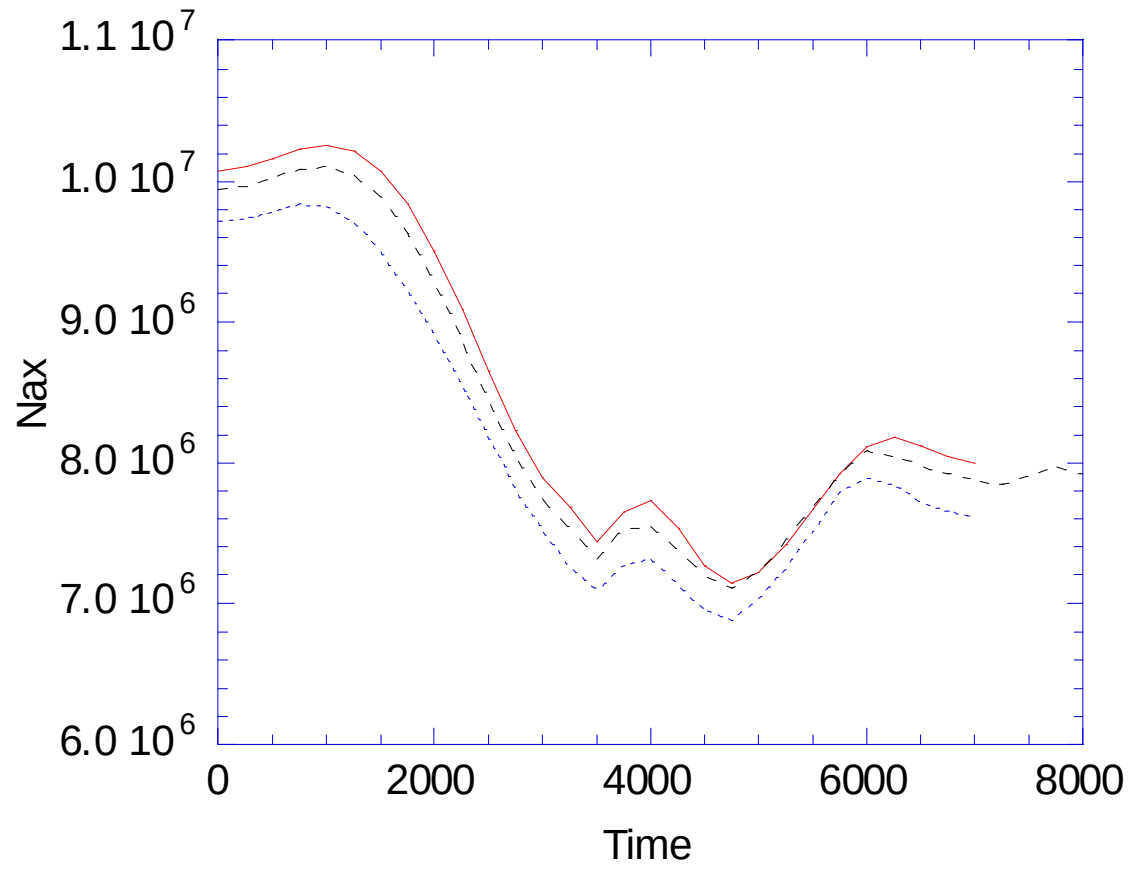

Figure 4: $N_{ax}$ of a circular loop as a function of time for $R_0$ = 2400, and λ = 0.004 (solid line), λ = 0.001 (dashed line), and λ = 0.00025 (dotted line).

## 2.2. Rotating loops

There exists a family of nonintersecting ("Kibble-Turok") [11,12,13,14] loops, which have been studied in the context of gauge strings. They are solutions of the Nambu-Goto equations of motion and all have non-zero angular momentum.

Intercommuting (self-intersection with reconnection) causes the loop sizes to shrink, and hence the average energy of radiated axions to increase and hence $N_{ax}$ to decrease. Intercommuting favors scenario B for these reasons. We picked the Kibble-Turok configuration as an initial condition to avoid intercommuting as much as possible, thus giving scenario A the best possible chance to get realized.

A common loop parameterization is given by

$$x(\sigma,t) = \frac{R}{2}\left\{(1-\alpha)\sin\sigma_- + \frac{1}{3}\alpha\sin 3\sigma_- + \sin\sigma_+\right\}$$
$$y(\sigma,t) = \frac{R}{2}\left\{-(1-\alpha)\cos\sigma_- - \frac{1}{3}\alpha\cos 3\sigma_- - \cos\psi\cos\sigma_+\right\}$$
$$z(\sigma,t) = \frac{R}{2}\left\{-2\sqrt{\alpha(1-\alpha)}\cos\sigma_- - \sin\psi\cos\sigma_+\right\} \qquad (9)$$

where $\sigma_\pm = (\sigma \pm t)/R$, and $\sigma \in (0, 2\pi R)$ is the length along the loop. For a significant subset [12,13,14] of the free parameters $\alpha \in (0,1)$, $\psi \in (-\pi, \pi)$ the loop never self-intersects. A noteworthy feature is the periodic appearance of cusps, where the string velocity momentarily reaches the speed of light. The motion of a gauge string is damped by emission of gravitational radiation and the loop diameter shrinks with time. The power is $P = \Gamma G \mu^2$, where $\mu$ is the energy per unit length, $G$ is the gravitational constant, and $\Gamma$ is a constant ~ 50 - 100 which depends on $\alpha, \psi$. Evidently, the power is independent of loop size, and the loop undergoes $1/(\Gamma G \mu) \approx 10^4$ oscillations in its lifetime. The radiation power spectrum was numerically determined [15,16] to be $P_k \propto 1/k^{4/3}$.

Axion strings are much more short-lived than gauge strings and radiate axions efficiently due to the strong topological coupling between string and field. No closed loop solutions are known however.

According to scenario A [7,15], the radiated power is $P \cong 50\ f_a^2$, independent of loop size, and the loop shrinks linearly with time. The expected number of oscillations per loop half-life is $\pi \ln(L/\delta)/50 \approx 4$ for $\ln(L/\delta) \approx 70$. In addition, $N_{ax}$ should increase by a factor $\ln(L/\delta)$. In scenario B on the other hand, $N_{ax}$ should remain approximately constant.

We performed numerous simulations of rotating loops on a 3D ($256^3$) lattice with periodic boundary conditions. Standard Fourier techniques were used for the spectrum analysis, and $N_{ax}$ was computed as a function of time using the dispersion relationship

$$\omega_{mnp} = \sqrt{2(3 - \cos k_m - \cos k_n - \cos k_p)}\,. \qquad (10)$$

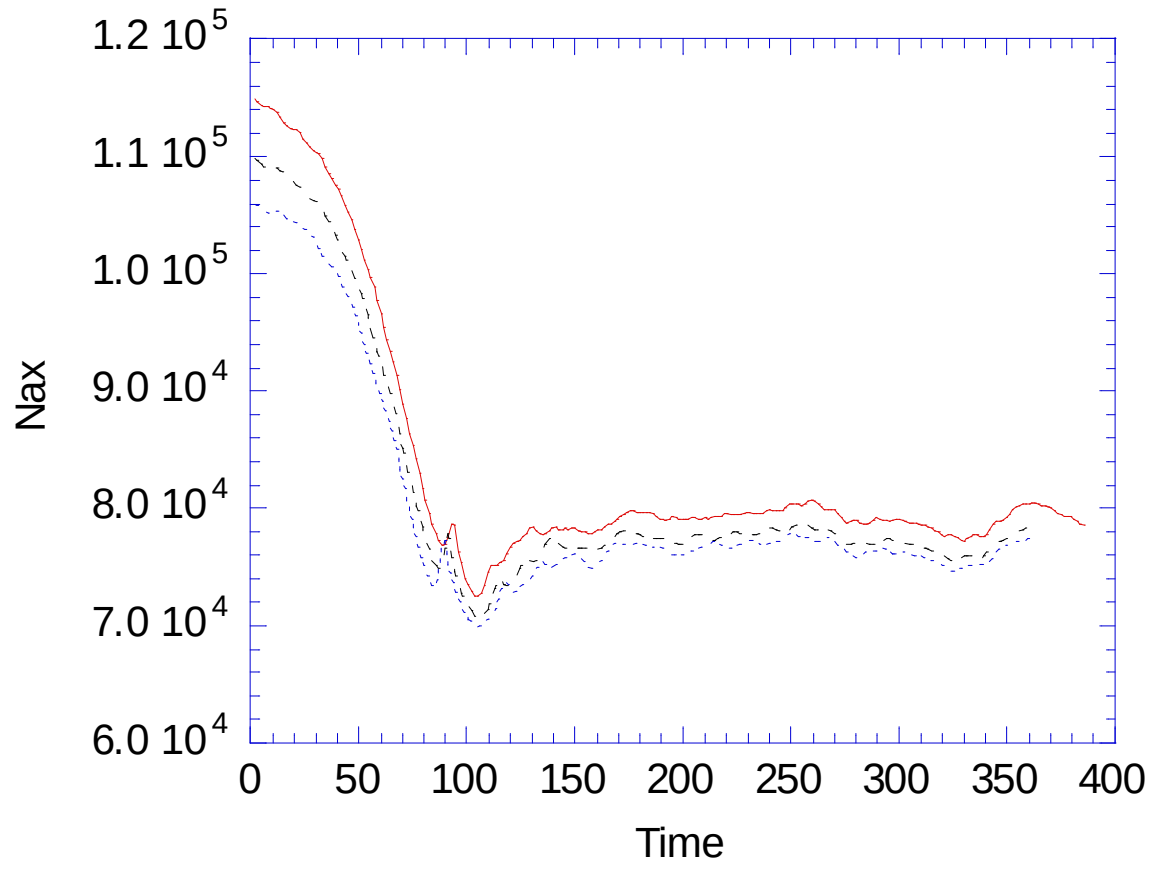

Figure 5: $N_{ax}$ of non-intersecting ("Kibble-Turok") loops as a function of time for α = 0.01, ϕ = 0, and λ = 0.2 (solid line), λ = 0.1 (dashed line), and λ = 0.0625 (dotted line). The lattice size is $256^3$ and $R$ = 72.

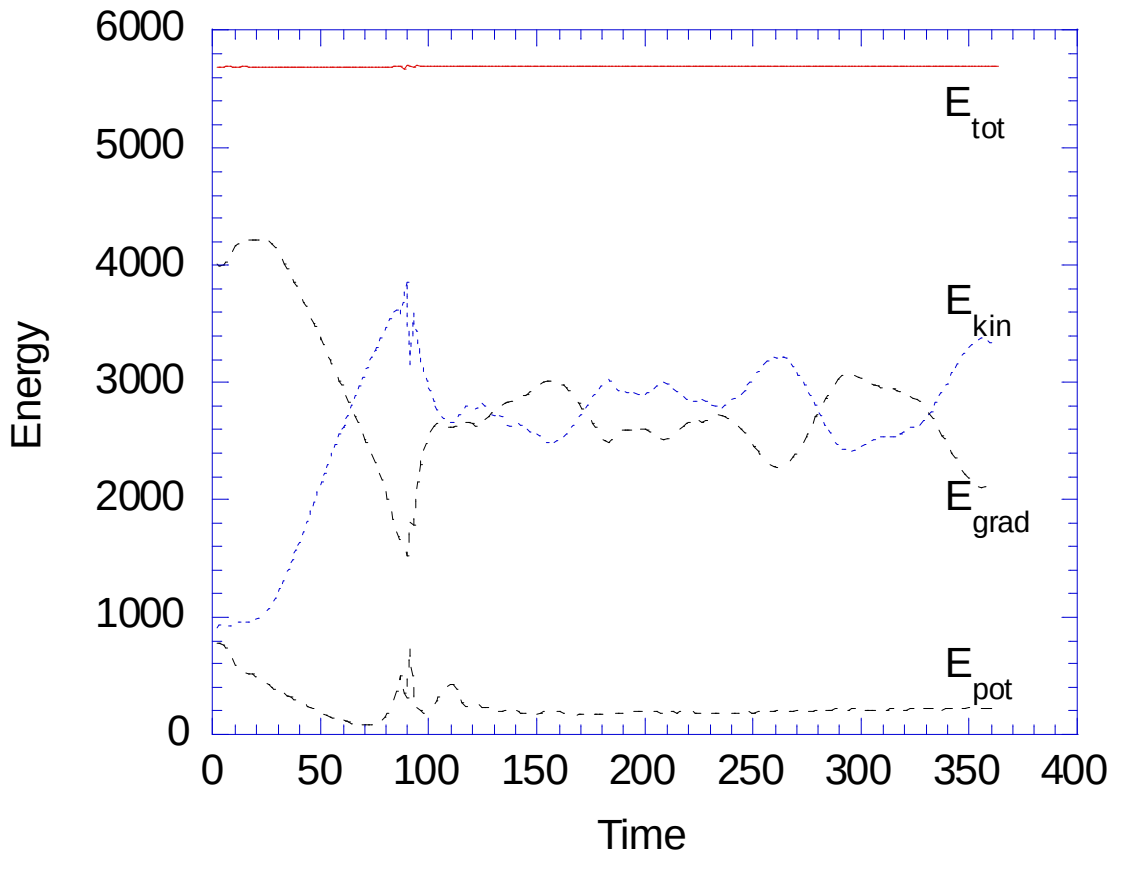

Figure 6: Energy of non-intersecting loop. Shown are total, gradient, kinetic, and potential energy as a function of time for $\alpha = 0.01$, $\psi = 0$, and $\lambda = 0.0625$.

Figure 5 shows $N_{ax}$ for various core sizes and constant $\alpha, \psi$. The behavior is very similar to that of a non-rotating circular loop with a reduction of ~ 25 %. Figure 6 depicts the energy of the collapsing loop. Clearly, the total energy is well conserved. A few percent of the loop energy is dissipated as massive radiation, shown here as $E_{pot}$.

## 2.3. String-antistring pairs

Lastly, we studied rotating, parallel string-antistring pairs, which can be thought of as cross-sections of loops with large eccentricity. The $2048^2$ lattice was initialized with the Abrikosov ansatz [17]

$$\phi(x,y) = \phi_1(x - x_1, y - y_1)\phi_2^*(x - x_2, y - y_2) \quad (11)$$

where $(x_1, y_1), (x_2, y_2)$ are the locations of string and antistring respectively. The fields were relaxed with periodic boundary conditions and the cores held fixed. The time derivative $\dot{\phi}(x,y)$ was obtained by a finite difference over a small time step. The attractive force per unit length between strings

$$F(\rho) = \frac{\partial V(\rho)}{\partial \rho} \approx 2\pi / \rho \quad (12)$$

is competing against the centrifugal force. In general, one expects a bound relativistic system to form, which decays by emission of axions and eventually annihilates.

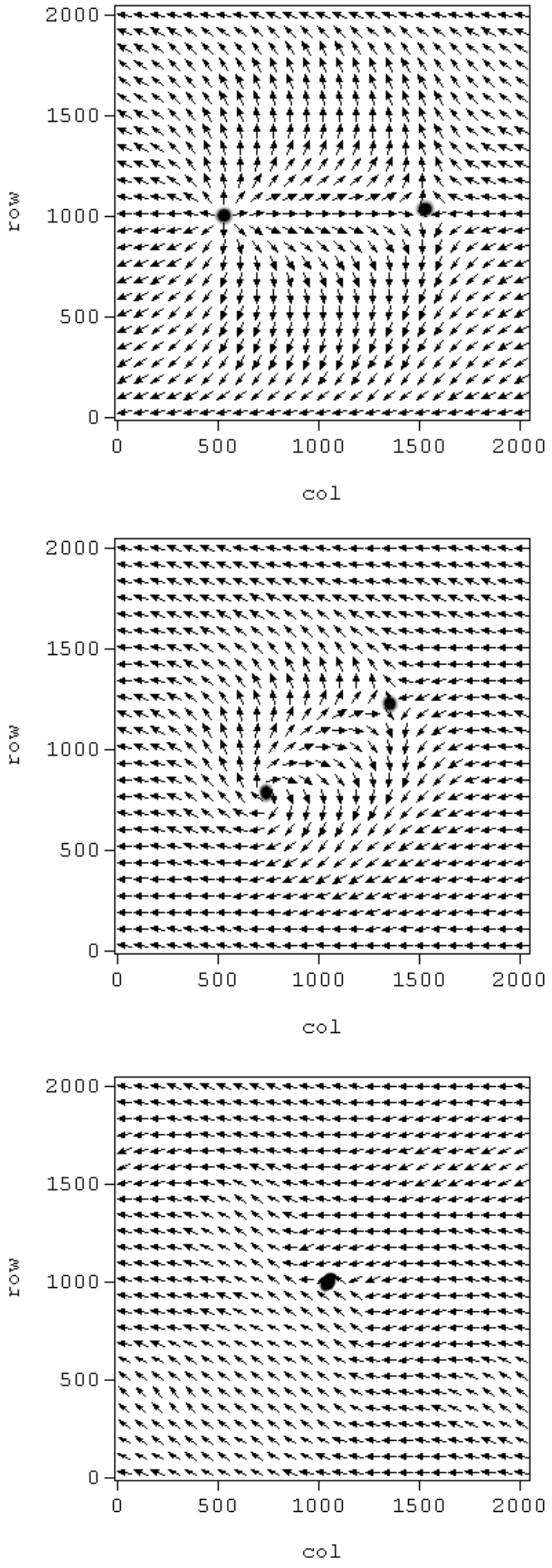

Figure 7: Collapsing string-antistring pair. The initial rotational velocity of the cores is 0.4 and $\lambda = 0.005$. The snapshots are for t = 40, 800, 1440.

Figure 7 shows the decay of a string-antistring system for an initial rotational velocity of 0.4. For the parameter range ($3 < \ln(\rho/\delta)$, $v_{rot} \leq 0.6$, $\rho \sim 500$), the system collapsed immediately in a time of order $\rho$.

The spectrum and $N_{ax}$ were computed as well and Figure 8 shows the evolution for several cases.

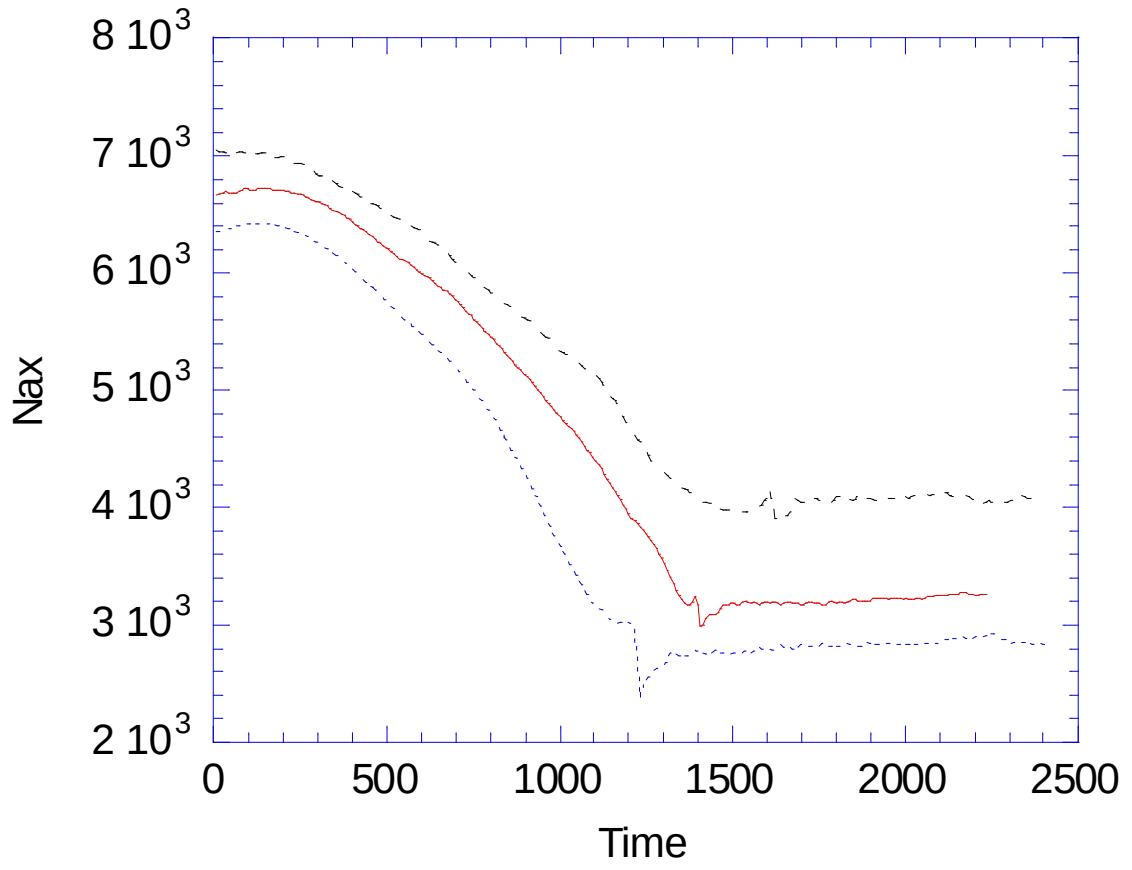

Figure 8: $N_{ax}$ of collapsing string-antistring pair. The initial rotational velocity is 0.6 (dashed), 0.4 (solid), 0.2 (dotted) and $\lambda = 0.005$.

## 3. CONCLUSION

Two different opinions exist about how axion strings decay. The first view (scenario A) states that the loops are moderately damped and oscillate several times during their lifetime. The instantaneous radiation spectrum is dominated by wavelengths of order the loop size, and $N_{ax}$ is predicted to increase during the decay by a factor $\ln(L/\delta)$.

Scenario B on the other hand proposes that the loops are critically damped, the radiation spectrum is $1/k$, and $N_{ax}$ is approximately constant. Our simulations clearly support scenario B since $N_{ax}$ decreases by about 20 % during the decay of axion loops for $\ln(L/\delta) \sim 6$. Moreover, the decrease of $N_{ax}$ which we observe during the decay of string loops, does not change with $\ln(L/\delta)$ over the range of $\ln(L/\delta)$ which we investigated.

## Acknowledgements

This work was performed under the auspices of the U.S. Department of Energy under Contract No. W-7405-Eng-48 at LLNL and under DE-FG05-86ER40272 at the University of Florida. We thank Livermore Computing for granting us generous access to their facilities.